\documentclass[journal,transmag]{IEEEtran}
\usepackage{graphicx}
\usepackage{graphics}
\usepackage{subfigure}
\usepackage{calligra}
\ifCLASSINFOpdf
\else
\fi
\begin{document}
\title{Quantum state complexity and the thermodynamic arrow of time}


%Montgomery Scott\IEEEauthorrefmark{3}, and
%Eldon Tyrell\IEEEauthorrefmark{4},~\IEEEmembership{Fellow,~IEEE}}
%\IEEEauthorblockA{\IEEEauthorrefmark{1}School of Electrical and Computer Engineering,
%Georgia Institute of Technology, Atlanta, GA 30332 USA}
%\IEEEauthorblockA{\IEEEauthorrefmark{2}Twentieth Century Fox, Springfield, USA}
%\IEEEauthorblockA{\IEEEauthorrefmark{3}Starfleet Academy, San Francisco, CA 96678 USA}
%\IEEEauthorblockA{\IEEEauthorrefmark{4}Tyrell Inc., 123 Replicant Street, Los Angeles, CA 90210 USA}% <-this % stops an unwanted space
%\thanks{Manuscript received December 1, 2012; revised August 26, 2015.
%Corresponding author: M. Shell (email: http://www.michaelshell.org/contact.html).}}

%\author{Xiao Dong, Jiasong Wu, Ling Zhou}
%
%\affiliation{Faculty of Computer Science and Engineering, Southeast University, Nanjing, China}
%
%\email{Xiao.Dong@seu.edu.cn; jswu@seu.edu.cn; Ling.Zhou@seu.edu.cn}

\author{\IEEEauthorblockN{Xiao Dong, Ling Zhou}
\IEEEauthorblockA{Faculty of Computer Science and Engineering, Southeast University, Nanjing, China}}

%{Shell \MakeLowercase{\textit{et al.}}: Bare Demo of IEEEtran.cls for IEEE Transactions on Magnetics Journals}



% for Transactions on Magnetics papers, we must declare the abstract and
% index terms PRIOR to the title within the \IEEEtitleabstractindextext
% IEEEtran command as these need to go into the title area created by
\IEEEtitleabstractindextext{%
\begin{abstract}
Why time is a one-way corridor? What's the origin of the arrow of time? We attribute the thermodynamic arrow of time to the direction of increasing quantum state complexity. Inspired by the work of Nielsen\cite{Nielsen_geometry2}, Susskind\cite{Susskind_ER_bridge_nowhere}\cite{Susskind_ER_bridge}\cite{Susskind2016The} and Micadei\cite{Micadei_timearrow}, we checked this hypothesis on both a simple two-qubit quantum system and a three-qubit system. The result shows that in the two-qubit system, the thermodynamic arrow of time always points in the direction of increasing quantum state complexity. For the three-qubit system, the heat flow pattern among its subsystems is closely correlated with the quantum state complexity of the subsystems. We propose that besides its impact on macroscopic spatial geometry\cite{Susskind_ER_bridge_nowhere}\cite{Susskind_ER_bridge}\cite{Susskind2016The}, quantum state complexity might also generate the thermodynamic arrow of time.

\end{abstract}

}

\maketitle

%\affiliation{}

%\tableofcontents

\IEEEdisplaynontitleabstractindextext

% \fi
\IEEEpeerreviewmaketitle

\section{Motivation}

The mysterious thermodynamic arrow of time has been debated for a long time. Physicists tried to address this problem from different points of view including entropy, quantum measurement, correlation, entanglement and complexity. But how about quantum state complexity?

Our work is inspired mainly by the work of Nielsen\cite{Nielsen_geometry2}, Susskind\cite{Susskind_ER_bridge_nowhere}\cite{Susskind_ER_bridge}\cite{Susskind2016The} and Micadei\cite{Micadei_timearrow}. We will scratch the road map of how the idea that quantum state complexity is connected with the thermodynamic arrow of time jumps into our mind by reviewing the main ideas of these papers.

\begin{itemize}
  \item \emph{Step 1. Quantum state complexity is physical}\\
  In \cite{Nielsen_geometry2} Nielsen proposed a geometric picture of quantum computation complexity. By defining a Riemannian metric on the manifold of unitary operation of quantum states, the complexity of any quantum computation algorithm in the quantum circuit model can be defined and we have the slogan \emph{quantum computation as free falling}.  Given the quantum computation complexity, the complexity of a quantum state $|\psi\rangle$ can also be defined as the minimal complexity of all the quantum circuits that can generate this state from a simple (separable) initial quantum state, for example $|00...0\rangle$. With this model, quantum state complexity is physically defined.

  Unfortunately it's generally difficult to compute the quantum state complexity. In \cite{Nielsen_geometry2} it's shown that the curvature of the constructed quantum computation manifold is almost non-positive so that the geodesic is not stable.

  \item \emph{Step 2. Quantum state complexity can lead to macroscopic spacetime effects}\\
  Maybe due to the complexity of computing quantum state complexity, the concept of quantum state complexity was ignored by physicists for a long time until Susskind saw it. He introduced the concept of quantum state complexity in his work on the geometry of black holes and wormholes. He claimed that quantum state complexity is related with the spatial volume and action\cite{Susskind_ER_bridge_nowhere}\cite{Susskind_ER_bridge}\cite{Brown2015Complexity}. Also in \cite{Susskind2016The} he indicated the complexity of a black hole as a quantum system increases monotonically for an exponentially long time. It's the increasement of quantum state complexity that keeps a black hole's horizon transparent and the complexity decreasing opaque state is extremely fragile. We can immediately see that this is really similar to the property of the thermodynamic arrow of time. It's very natural to check how the quantum state complexity is related with time.

  \item \emph{Step 3. Testify the relation between quantum state complexity and the arrow of time}\\
  The recent experimental work of Micadei\cite{Micadei_timearrow} showed the reversal of the thermodynamic arrow of time using quantum correlation. In their discussion, they attribute the observed reversal of the arrow of time to the quantum correlation just as in \cite{Partovi2008Entanglement} \cite{Jennings2010Entanglement}. Their work inspired us to verify our idea on the relation of quantum state complexity of time, because they used a simple two-qubit system to achieve their task and it's not that difficult to compute the complexity of a two qubit system.

  Our idea is simple. If we can show that the change of the direction of the arrow of time is correlated with the change of the quantum state complexity on such a two-qubit system, then we have at least the first concrete example that the thermodynamic arrow of time may stem from quantum state complexity.

\end{itemize}

\section{The thermodynamic arrow of time and quantum state complexity on a 2-qubit system}

\subsection{System setup}

Following the setup of \cite{Micadei_timearrow}, we focus on a 2-qubit system (A,B) with an initial state of the form,
\begin{equation}\label{eq_1}
  \rho_{AB}^{0}=\rho_{A}^{0}\otimes\rho_{B}^{0}+\chi_{AB},
\end{equation}
where $\chi_{AB}=\alpha|01\rangle\langle10|+\alpha^{*}|10\rangle\langle01|$ is the correlation term with a proper $\alpha$  to ensure the positivity of $\rho_{AB}$. $\rho_{i}^{0}=exp(-\beta_{i}H_{i})/Z_{i}$ is a thermal state at the inverse temperature $\beta_{i}$, $i=(A,B)$ for qubit A and B respectively, and $Z_i=Tr(exp(-\beta_{i}H_{i}))$ is the partition function. The state $|0\rangle$ and $|1\rangle$ are the ground and the excited eigenstates of the Hamiltonian $H_{i}=h\nu_{0}(1-\sigma_z^{i})/2$. For simplicity, we set $h\nu_{0}=1$ so that $H_{i}=(1-\sigma_{z}^{i})/2$. Also the system will evolve under an effective interaction Hamiltonian $H_{AB}^{eff}=(\pi/2)(\sigma_{x}^{A}\sigma_{y}^{B}-\sigma_{y}^{A}\sigma_{x}^{B})$. In such a system, no work is performed and the heat absorbed by one qubit is given by its internal energy variation along the dynamics so that $Q_{i}=\Delta E_{i}$ with $E_{i}=Tr_{i}H_{i}\rho_{i}$. For more details of the system setup, please refer to \cite{Micadei_timearrow}.

We check the thermodynamic arrow of time under three scenarios by simulations.
\begin{enumerate}
  \item $\alpha=0$: This is the case that there is no correlation in the initial state, so the arrow of time will be normal when the system start to evolve.
  \item $\alpha=0.1$: In this setup there exists \emph{normal} correlation between A and B and this can lead to a reversal of the thermodynamic arrow of time as illustrated in \cite{Micadei_timearrow}.
  \item $\alpha=0.1e^{i\pi/2}$: According to \cite{Micadei_timearrow}, $\chi_{AB}$ can not commute with the thermalization Hamiltonian $H_{AB}^{eff}$ so that the reversal of the arrow of time will not occur.
\end{enumerate}

\subsection{Quantum state complexity}
It should be emphasized that the Nielsen's version of quantum state complexity can not be directly implemented here since we are now working with mixted states. We need to generalize the definition of quantum state complexity.

We define \textbf{\emph{the state complexity of a quantum state $|\psi\rangle$ as the minimal length of all the curves that connect $|\psi\rangle$ to any state in the set of simple (zero complexity) states by unitary operations}}. In the language of Nielsen's quantum circuit complexity, this is the complexity of the most efficient quantum circuit that can generate $|\psi\rangle$ from a simple quantum state by unitary operations. This definition is very similar to the geometrical entanglement measure of an entangled state, which is defined as its minimal distance to the set of separable states.

But what's the set of zero complexity quantum states? And under which metric the distance is defined?

\begin{itemize}
  \item \emph{Zero complexity quantum states:} For a n-qubit quantum system, the zero complexity pure state set is defined as all the computational basis states. For example, a 2-qubit system, the set of zero complexity states is $\{|00\rangle , |01\rangle , |10\rangle , |11\rangle \}$. For mixed states, the zero complexity states are all the states with diagonal density matrix. Since we define the complexity on unitary operations, for a given mixed quantum state $\rho$ with a certain spectrum, it can only be generated from a zero complexity state with the same spectrum. For example, a 2-qubit mixed state with a spectrum $\{\lambda_1,\lambda_2,\lambda_3,\lambda_4\}$ can only be generated from zero complexity quantum states as $\lambda_{p(1)}|00\rangle+\lambda_{p(2)}|01\rangle+\lambda_{p(3)}|10\rangle+\lambda_{p(4)}|11\rangle$, where $\{\lambda_{p(1)},\lambda_{p(2)},\lambda_{p(3)},\lambda_{p(4)}\}$ is a permutation of $\{\lambda_{1},\lambda_{2},\lambda_{3},\lambda_{4}\}$. Obviously for a mixed state with a non-degenerated spectrum, it can be generated from 24 different zero complexity quantum states by unitary operations. And the state complexity of $\rho$ is its minimal distance to these 24 states. It should be noted that the so-called zero complexity mixed states here are not really low complexity states since it's generally difficult to generate such kind of mixed states, for example by tracing out a subsystem a a larger pure state system. But among all the mixed states with a given spectrum, they can be regarded as the states with the smallest complexity since they can be described by the smallest number of parameters. So we use them as a baseline to compute the relative complexity of a mixed state.

  \item \emph{Distance metric:} The length of a curve need to be defined under a certain distance metric. In fact we have different choices. Typical metrics include Nilsen's Riemannian metric\cite{Nielsen_geometry2}, Bures distance\cite{Braunstein2008Geometry} and the dynamic distance measure\cite{Heydari_dynamicdiatance}. In this work we use the simple Bures distance.
\end{itemize}

\subsection{Connecting the arrow of time with quantum state complexity}
We try to find the correlation between the thermodynamic arrow of time and the quantum state complexity by simulating the quantum evolution of the above described 2-qubit system under the effective thermodynamic interaction $H_{AB}^{eff}$. The state of the system is then given by $\rho_{AB}^{t}=U(t)\rho_{AB}^{0}U(t)^{+}$ and $U(t)=exp(-itH_{AB}^{eff})$.

We set the initial inverse temperature $\beta_{A}=1.0$ and $\beta_{B}=2.0$. We then check how the internal energies of the two subsystems vary with time in the above mentioned three scenarios ($\alpha=0,0.1,0.1i$).

At the same time, we compute the state complexity of $\rho_{AB}^{t}$ by Bures distance $D_{B}$, which is given explicitly for 2-qubit systems by

\begin{eqnarray}
  D_B^2(\rho_1,\rho_2) &=& Tr\rho_1+Tr\rho_2-2\sqrt{F}(\rho_1,\rho_2) \\
  \sqrt{F}(\rho_1,\rho_2) &=& Tr\sqrt{\sqrt{\rho_2}\rho_1\sqrt{\rho_2}}
\end{eqnarray}
where $\sqrt{\rho_2}$ is the unique positive square root of $\rho_2$.

The state complexity of $\rho_{AB}^{t}$ is defined by the minimal Bures distance between $\rho_{AB}^{t}$ and all the 24 0-complexity mixed states with the same spectrum of $\rho_{AB}^{t}$.

\begin{figure}
  \centering
  \includegraphics[width=8cm]{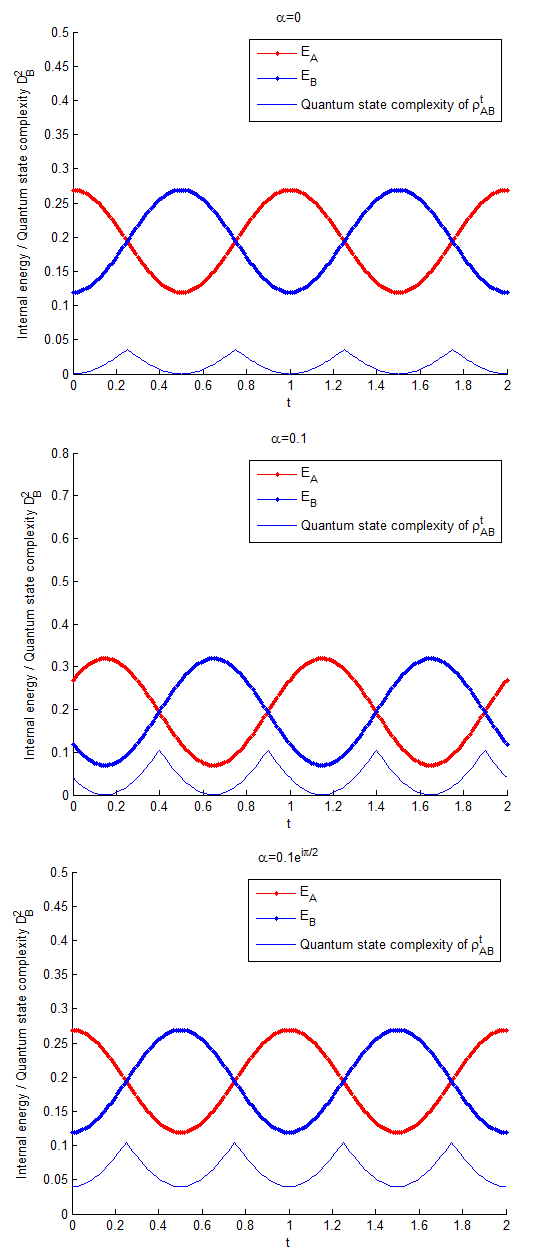}
  \caption{Simulation of the 2-qubit system under three scenarios given by $\alpha=0,0.1$ and $0.1i$. In each case, we give the internal energies of the two subsystems as $E_A,E_B$ and the quantum state complexity of $\rho_{AB}^{t}$ is computed based on Bures distance and our definition of quantum state complexity. The thermodynamic arrow of time is strictly consistent with the change of quantum state complexity so that the arrow of time corresponds to the increasing quantum state complexity. Also for the case of $\alpha=0$ and $\alpha=0.1i$, we have the same time flow pattern and similar state complexity patterns. }\label{fig1}
\end{figure}

The result is shown in Fig. \ref{fig1}. It can be observed that in all the three scenarios, the thermodynamic arrow of time always points in the direction of increasing quantum state complexity. This is to say, when the quantum state complexity is decreasing, the arrow of time is reversed so the heat flows from the low temperature (internal energy) subsystem to the high temperature subsystem. And when the state complexity is increasing, we see a normal thermodynamic arrow of time.

\begin{figure}
  \centering
  \includegraphics[width=8cm]{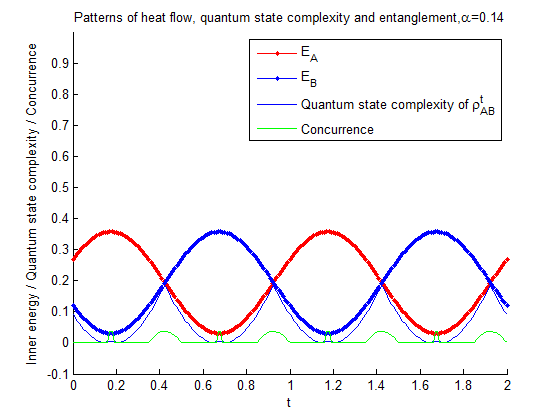}
  \caption{Relations of the heat flow pattern, quantum state complexity and entanglement of formation (concurrence) for $\alpha=0.14$. Still the thermodynamic arrow of time matches the quantum state complexity perfectly. But there is no correspondence between the time of arrow and the concurrence.}\label{fig6}
\end{figure}

In order to check the relation between the arrow of time and the entanglement, we also compute the concurrence $C(\rho_{AB}^t)$ of $\rho_{AB}^t$, which is connected with the entanglement of formation by $EoF=h(\frac{1}{2}+\frac{1}{3}\sqrt{1-(C(\rho_{AB}^2))})$ with $h(x)=-xlog(x)-(1-x)log(1-x)$ \cite{Streltsov2010Linking}. We set $\alpha=0.14$ (this value is selected so that entanglement emerges in the system and also $\rho_{AB}$ is positive) and the result is given in Fig. \ref{fig6}. We can see a perfect matching between the arrow of time and the quantum state complexity, but there is no correspondence between the arrow of time and the entanglement. This is a sign that quantum state complexity is more suitable to generate the arrow of time than entanglement.

\subsection{A three-qubit system}
To further verify the relationship between quantum state complexity and the arrow of time, we also checked a three-qubit quantum system proposed in \cite{Jennings2010Entanglement}. The system is initially in a mixed state $\rho_{ABC}$, which has the property that $\rho_{AB}=\rho_A(T_A)\otimes\rho_B(T_B)$ and $\rho_{BC}=\rho_B(T_B)\otimes\rho_C(T_C)$ for thermal marginals $\rho_i(T_i)$,$i=(A,B,C)$ and $T_A<T_B<T_C$. Here all qubits have the same Hamiltonian $H_i=(I+\sigma_z^i)/2$. But the initial $\rho_{AC}$ is more complex and given by

\begin{eqnarray*}
   \rho_{AC}&=&\frac{1}{2}(\gamma+\lambda_C-\lambda_A)|10\rangle\langle10|+(\gamma-\lambda_C+\lambda_A)|01\rangle\langle01|\\
   &+&\sqrt{\gamma^2-(\lambda_C-\lambda_A)^2}(|10\rangle\langle10|+|01\rangle\langle10|) \\
   &+&(\lambda_A+\lambda_C-\gamma)|00\rangle\langle00|+(2-\lambda_A-\lambda_C-\gamma)|11\rangle\langle11| \\
   &+&\sqrt{(\lambda_A+\lambda_C-\gamma)(2-\lambda_A-\lambda_C-\gamma)}(|00\rangle\langle11|+|11\rangle\langle00|) \\
\end{eqnarray*}

In this work we set $T_A=4, T_B=2, T_C=1, \lambda_A=0.15, \lambda_C=0.3, \gamma=0.4$.

The total state is now $\rho_{ABC}=\rho_{AC}\otimes\rho_B$ and the interaction Hamiltonians are $H_{AB}=(\sigma_X^A\sigma_Y^B-\sigma_Y^A\sigma_X^B)/2$ and $H_{BC}=(\sigma_X^B\sigma_Y^C-\sigma_Y^B\sigma_X^C)/2$. Generally the system will evolve under the unitary $U(t,s,\tau)=exp[-i(tH_{AB}+sH_{BC})\tau)]$, where $s,t$ are the strengths of the interactions and $\tau$ stands for time.

 As illustrated in \cite{Jennings2010Entanglement}, the system shows a complex heat flow pattern depending on the variables $s,t$. Here we will check how the pattern of heat flow, which is related with the thermodynamic arrow of time, is related with the quantum state complexity.

\begin{figure}
  \centering
  \includegraphics[width=9cm]{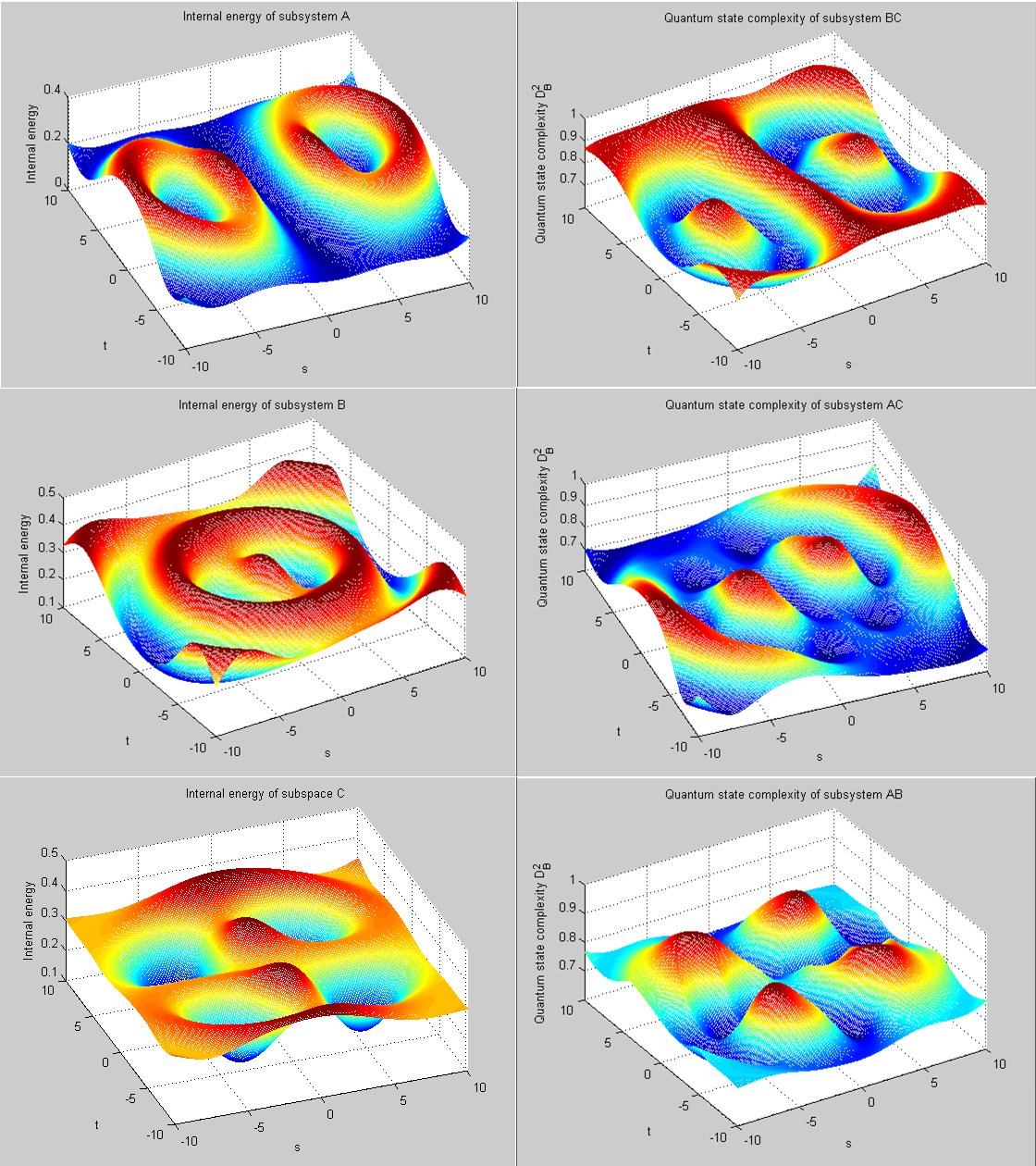}
  \caption{Internal energy and quantum state complexity of the three qubit system. We can clearly see there is a high similarity between the pattern of internal energies (heat flow) and state complexities of subsystems. This is a strong sign that the thermodynamic arrow of time is closely related with the quantum state complexity.}\label{fig4}
\end{figure}

We first fixed $\tau=1$ and set $s\in [-10,10]$, $t\in [-10,10]$. This means we vary the interaction strength and evolve the system for a fixed time period. We then compute the internal energy of each qubit as $E_i=Tr(\rho_iH_i)$ and the quantum state complexity of $\rho_{AB},\rho_{BC},\rho_{AC}$ using Bures distance. The result is given in Fig. \ref{fig4}.

\begin{figure}
  \centering
  \includegraphics[width=9cm]{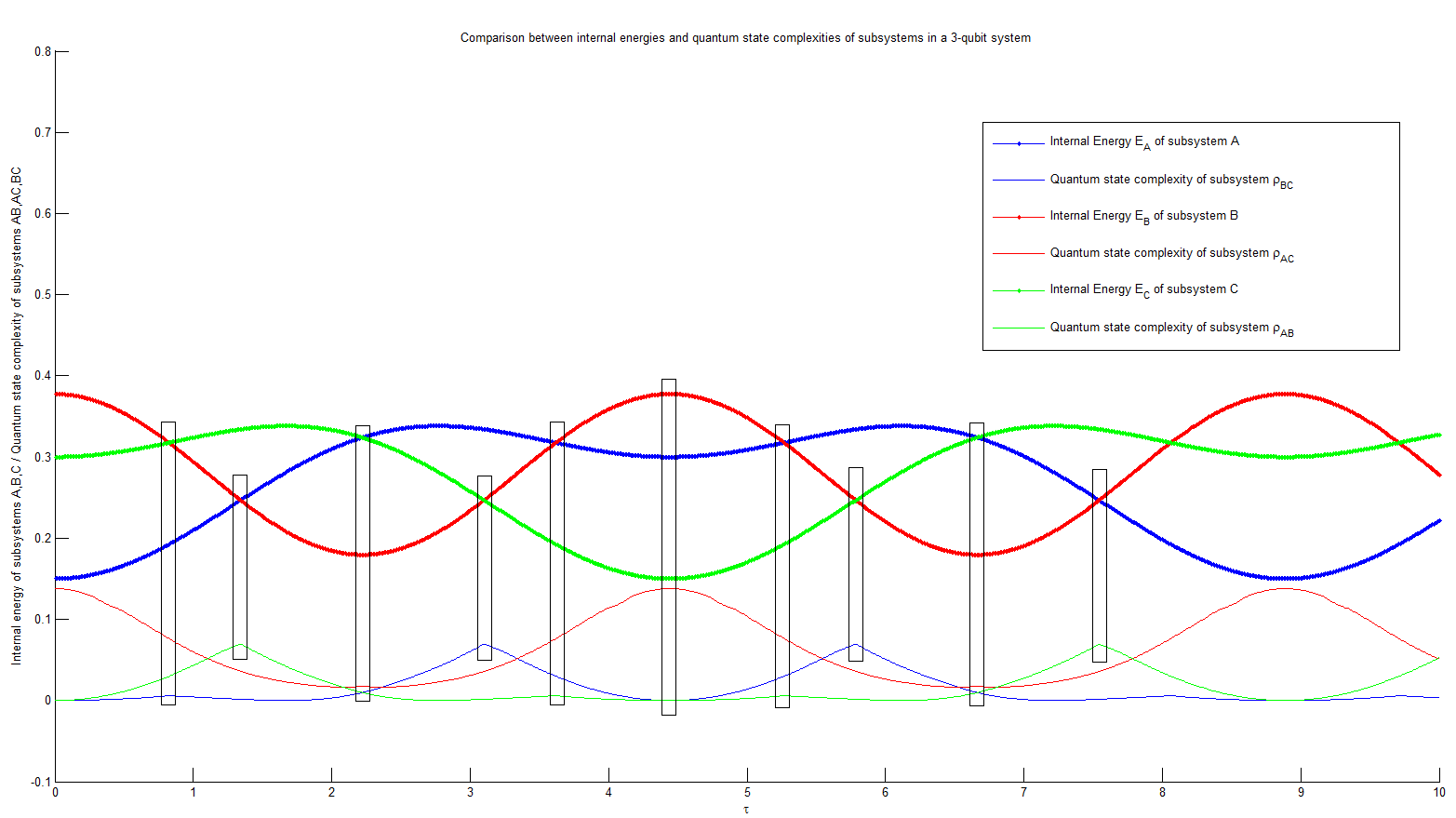}
  \caption{Internal energy and quantum state complexity of the three qubit system for $s=t=1$ and $\tau\in[0,10]$. Still the strong correlation between heat flow patterns and quantum state complexity patterns is observed. Key points of these patterns are highlighted by the rectangles. All the changing points in the heat flow pattern and the state complexity pattern are synchronous.}\label{fig5}
\end{figure}

As mentioned in \cite{Jennings2010Entanglement}, setting either $s$ or $t$ to zero is trivial since this returns to the 2-qubit case, which has been checked above. We are now interested in the situation that both $s$ and $t$ are nonzero so that the interactions between AB and BC are both turned on. To further explore the details, we check a typical situation where $s=t=1$ and we let the system evolve for a time interval $\tau\in[-10,10]$.  The result is shown in Fig. \ref{fig5}.

Now the heat flow pattern is among three subsystems, so it's difficult to define the thermodynamic arrow of time and connect it with the state complexity as in the 2-qubit case. But we can see clearly that the heat flow pattern and the state complexity pattern are very similar as shown in Fig. \ref{fig4}. In Fig. \ref{fig5} we see the changes in the state complexity patterns and the internal energy patterns are synchronous. This synchronization is a strong sign for the close relation between them.

\section{Conclusions}
The reversal of the thermodynamic arrow of time has been addressed as an emergent  phenomenon of the correlation or entanglement patterns between subsystems.
We propose to understand the thermodynamic arrow of time as a result of the change of the quantum state complexity.

We verified our hypothesis on both a simple 2-qubit system and a more complicated 3-qubit system. In both cases we see strong correlation between the arrow of time and the quantum state complexity.

If we go back to Susskind, he said that for a large quantum system such as a black hole, statistically its quantum state complexity increases linearly for an exponentially long time before the complexity saturates. This stable and linear complexity increasing pattern seems a perfect picture for Newton's smooth time flow.  Combing this work with Susskind's idea that complexity is related with spatial volume, we may say quantum state complexity can have microscopic effect on spacetime geometry.

Note: After finishing this work, we noticed the work of Baumeler\cite{Baumeler2016Causality}. They proposed to use the intrinsic physically motivated measure to describe the randomness of a string of bits, which is closely related with Kolmogorov complexity. In fact our definition of quantum state complexity is exactly a Kolmogorov complexity of a quantum state, which is defined as the length of the shortest quantum algorithm to generate this state.

\bibliographystyle{unsrt}

\bibliography{timearrow}



%\vfill


\end{document}